\documentclass[aps,prl,twocolumn,groupedaddress]{revtex4}
\usepackage{amsmath}
\usepackage{graphicx}

\begin{document}

%\preprint{}

\hspace{5.2in}\mbox{FERMILAB-PUB-06-268-E}

\title{Limits on anomalous trilinear gauge couplings from 
$WW \to e^+e^-$, $WW \to e^\pm \mu^\mp$, and $WW \to \mu^+\mu^-$ events
from $p\bar{p}$ collisions at $\sqrt{s}=1.96$ TeV
}


%\author{}
%\thanks{}
%\affiliation{}
% LIST_OF_AUTHORS_R2.TEX                 7/19/06            
%
\author{                                                                      
%% names begin here                                                           
V.M.~Abazov,$^{36}$                                                           
B.~Abbott,$^{76}$                                                             
M.~Abolins,$^{66}$                                                            
B.S.~Acharya,$^{29}$                                                          
M.~Adams,$^{52}$                                                              
T.~Adams,$^{50}$                                                              
M.~Agelou,$^{18}$                                                             
S.H.~Ahn,$^{31}$                                                              
M.~Ahsan,$^{60}$                                                              
G.D.~Alexeev,$^{36}$
% ...
\\                                                                            
\vskip 0.30cm                                                                 
\centerline{(D\O\ Collaboration)}                                             
\vskip 0.30cm                                                                 
}                                                                             
\affiliation{                                                                 
\centerline{$^{1}$Universidad de Buenos Aires, Buenos Aires, Argentina}       
\centerline{$^{2}$LAFEX, Centro Brasileiro de Pesquisas F{\'\i}sicas,         
                  Rio de Janeiro, Brazil}                                     
\centerline{$^{3}$Universidade do Estado do Rio de Janeiro,                   
                  Rio de Janeiro, Brazil}                                     
\centerline{$^{4}$Instituto de F\'{\i}sica Te\'orica, Universidade            
                  Estadual Paulista, S\~ao Paulo, Brazil}                     
\centerline{$^{5}$University of Alberta, Edmonton, Alberta, Canada,           
                  Simon Fraser University, Burnaby, British Columbia, Canada,}
\centerline{York University, Toronto, Ontario, Canada, and                    
                  McGill University, Montreal, Quebec, Canada}                
\centerline{$^{6}$Institute of High Energy Physics, Beijing,                  
                  People's Republic of China}                                 
\centerline{$^{7}$University of Science and Technology of China, Hefei,       
                  People's Republic of China}                                 
\centerline{$^{8}$Universidad de los Andes, Bogot\'{a}, Colombia}             
\centerline{$^{9}$Center for Particle Physics, Charles University,            
                  Prague, Czech Republic}                                     
\centerline{$^{10}$Czech Technical University, Prague, Czech Republic}        
\centerline{$^{11}$Center for Particle Physics, Institute of Physics,         
                   Academy of Sciences of the Czech Republic,                 
                   Prague, Czech Republic}                                    
\centerline{$^{12}$Universidad San Francisco de Quito, Quito, Ecuador}        
\centerline{$^{13}$Laboratoire de Physique Corpusculaire, IN2P3-CNRS,         
                   Universit\'e Blaise Pascal, Clermont-Ferrand, France}      
\centerline{$^{14}$Laboratoire de Physique Subatomique et de Cosmologie,      
                   IN2P3-CNRS, Universite de Grenoble 1, Grenoble, France}    
\centerline{$^{15}$CPPM, IN2P3-CNRS, Universit\'e de la M\'editerran\'ee,     
                   Marseille, France}                                         
\centerline{$^{16}$IN2P3-CNRS, Laboratoire de l'Acc\'el\'erateur              
                   Lin\'eaire, Orsay, France}                                 
\centerline{$^{17}$LPNHE, IN2P3-CNRS, Universit\'es Paris VI and VII,         
                   Paris, France}                                             
\centerline{$^{18}$DAPNIA/Service de Physique des Particules, CEA, Saclay,    
                   France}                                                    
\centerline{$^{19}$IPHC, IN2P3-CNRS, Universit\'e Louis Pasteur, Strasbourg,  
                    France, and Universit\'e de Haute Alsace,                 
                    Mulhouse, France}                                         
\centerline{$^{20}$Institut de Physique Nucl\'eaire de Lyon, IN2P3-CNRS,      
                   Universit\'e Claude Bernard, Villeurbanne, France}         
\centerline{$^{21}$III. Physikalisches Institut A, RWTH Aachen,               
                   Aachen, Germany}                                           
\centerline{$^{22}$Physikalisches Institut, Universit{\"a}t Bonn,             
                   Bonn, Germany}                                             
\centerline{$^{23}$Physikalisches Institut, Universit{\"a}t Freiburg,         
                   Freiburg, Germany}                                         
\centerline{$^{24}$Institut f{\"u}r Physik, Universit{\"a}t Mainz,            
                   Mainz, Germany}                                            
\centerline{$^{25}$Ludwig-Maximilians-Universit{\"a}t M{\"u}nchen,            
                   M{\"u}nchen, Germany}                                      
\centerline{$^{26}$Fachbereich Physik, University of Wuppertal,               
                   Wuppertal, Germany}                                        
\centerline{$^{27}$Panjab University, Chandigarh, India}                      
\centerline{$^{28}$Delhi University, Delhi, India}                            
\centerline{$^{29}$Tata Institute of Fundamental Research, Mumbai, India}     
\centerline{$^{30}$University College Dublin, Dublin, Ireland}                
\centerline{$^{31}$Korea Detector Laboratory, Korea University,               
                   Seoul, Korea}                                              
\centerline{$^{32}$SungKyunKwan University, Suwon, Korea}                     
\centerline{$^{33}$CINVESTAV, Mexico City, Mexico}                            
\centerline{$^{34}$FOM-Institute NIKHEF and University of                     
                   Amsterdam/NIKHEF, Amsterdam, The Netherlands}              
\centerline{$^{35}$Radboud University Nijmegen/NIKHEF, Nijmegen, The          
                  Netherlands}                                                
\centerline{$^{36}$Joint Institute for Nuclear Research, Dubna, Russia}       
\centerline{$^{37}$Institute for Theoretical and Experimental Physics,        
                   Moscow, Russia}                                            
\centerline{$^{38}$Moscow State University, Moscow, Russia}                   
\centerline{$^{39}$Institute for High Energy Physics, Protvino, Russia}       
\centerline{$^{40}$Petersburg Nuclear Physics Institute,                      
                   St. Petersburg, Russia}                                    
\centerline{$^{41}$Lund University, Lund, Sweden, Royal Institute of          
                   Technology and Stockholm University, Stockholm,            
                   Sweden, and}                                               
\centerline{Uppsala University, Uppsala, Sweden}                              
\centerline{$^{42}$Physik Institut der Universit{\"a}t Z{\"u}rich,            
                   Z{\"u}rich, Switzerland}                                   
\centerline{$^{43}$Lancaster University, Lancaster, United Kingdom}           
\centerline{$^{44}$Imperial College, London, United Kingdom}                  
\centerline{$^{45}$University of Manchester, Manchester, United Kingdom}      
\centerline{$^{46}$University of Arizona, Tucson, Arizona 85721, USA}         
\centerline{$^{47}$Lawrence Berkeley National Laboratory and University of    
                   California, Berkeley, California 94720, USA}               
\centerline{$^{48}$California State University, Fresno, California 93740, USA}
\centerline{$^{49}$University of California, Riverside, California 92521, USA}
\centerline{$^{50}$Florida State University, Tallahassee, Florida 32306, USA} 
\centerline{$^{51}$Fermi National Accelerator Laboratory,                     
            Batavia, Illinois 60510, USA}                                     
\centerline{$^{52}$University of Illinois at Chicago,                         
            Chicago, Illinois 60607, USA}                                     
\centerline{$^{53}$Northern Illinois University, DeKalb, Illinois 60115, USA} 
\centerline{$^{54}$Northwestern University, Evanston, Illinois 60208, USA}    
\centerline{$^{55}$Indiana University, Bloomington, Indiana 47405, USA}       
\centerline{$^{56}$University of Notre Dame, Notre Dame, Indiana 46556, USA}  
\centerline{$^{57}$Purdue University Calumet, Hammond, Indiana 46323, USA}    
\centerline{$^{58}$Iowa State University, Ames, Iowa 50011, USA}              
\centerline{$^{59}$University of Kansas, Lawrence, Kansas 66045, USA}         
\centerline{$^{60}$Kansas State University, Manhattan, Kansas 66506, USA}     
\centerline{$^{61}$Louisiana Tech University, Ruston, Louisiana 71272, USA}   
\centerline{$^{62}$University of Maryland, College Park, Maryland 20742, USA} 
\centerline{$^{63}$Boston University, Boston, Massachusetts 02215, USA}       
\centerline{$^{64}$Northeastern University, Boston, Massachusetts 02115, USA} 
\centerline{$^{65}$University of Michigan, Ann Arbor, Michigan 48109, USA}    
\centerline{$^{66}$Michigan State University,                                 
            East Lansing, Michigan 48824, USA}                                
\centerline{$^{67}$University of Mississippi,                                 
            University, Mississippi 38677, USA}                               
\centerline{$^{68}$University of Nebraska, Lincoln, Nebraska 68588, USA}      
\centerline{$^{69}$Princeton University, Princeton, New Jersey 08544, USA}    
\centerline{$^{70}$State University of New York, Buffalo, New York 14260, USA}
\centerline{$^{71}$Columbia University, New York, New York 10027, USA}        
\centerline{$^{72}$University of Rochester, Rochester, New York 14627, USA}   
\centerline{$^{73}$State University of New York,                              
            Stony Brook, New York 11794, USA}                                 
\centerline{$^{74}$Brookhaven National Laboratory, Upton, New York 11973, USA}
\centerline{$^{75}$Langston University, Langston, Oklahoma 73050, USA}        
\centerline{$^{76}$University of Oklahoma, Norman, Oklahoma 73019, USA}       
\centerline{$^{77}$Oklahoma State University, Stillwater, Oklahoma 74078, USA}
\centerline{$^{78}$Brown University, Providence, Rhode Island 02912, USA}     
\centerline{$^{79}$University of Texas, Arlington, Texas 76019, USA}          
\centerline{$^{80}$Southern Methodist University, Dallas, Texas 75275, USA}   
\centerline{$^{81}$Rice University, Houston, Texas 77005, USA}                
\centerline{$^{82}$University of Virginia, Charlottesville,                   
            Virginia 22901, USA}                                              
\centerline{$^{83}$University of Washington, Seattle, Washington 98195, USA}  
}                                                                             
%end                                                                          

%\collaboration{D\O\ Collaboration}

%\date{\today}
\date{August 4, 2006}

\begin{abstract}
Limits are set on anomalous $WW\gamma$ and $WWZ$ trilinear gauge couplings using
$W^+W^-\to e^+\nu_ee^-\bar{\nu}_e$, $W^+W^- \to e^\pm\nu_e\mu^\mp\nu_\mu$, 
and $W^+W^- \to \mu^+\nu_\mu\mu^-\bar{\nu}_\mu$ events.  The data set was collected
by the Run II D\O~detector at the Fermilab Tevatron Collider and corresponds to
approximately $250 \text{ pb}^{-1}$ of integrated luminosity at $\sqrt{s}=1.96$ TeV.
Under the assumption that the $WW\gamma$ couplings are equal to the $WWZ$ 
couplings and using a form factor scale of $\Lambda = 2.0\text{ TeV}$, the combined 95\%
C.L. one-dimensional coupling limits from all three channels 
are $-0.32 < \Delta\kappa < 0.45$ and $-0.29 < \lambda < 0.30$.
\end{abstract}

\pacs{}
%\keywords{}

\maketitle

% References should be done using the  \cite, \ref, and \label commands
%\section{}
%\subsection{}
%\subsubsection{}

%
% INTRODUCTION SECTION
%
%\section{Introduction \label{sec-intro}}

Within the standard model (SM), interactions between the bosons of the
electroweak interaction are entirely determined by the gauge
symmetry. Any deviation from the SM couplings is
therefore evidence of new physics.

The most general Lorentz invariant effective Lagrangian which describes the
triple gauge couplings has fourteen independent coupling parameters,
seven for each of the $WW\gamma$ and $WWZ$ vertices \cite{Hagiwara:1989mx}.
With the assumption of
electromagnetic gauge invariance and $C$ and $P$ conservation, the
number of independent couplings is reduced to five, and the
Lagrangian takes this form: 

\begin{equation}
\label{eq-lag}
\begin{split}
\frac{\mathcal{L}_{WWV}}{g_{WWV}} & = ig^V_1(W^\dagger_{\mu\nu}W^{\mu}V^{\nu} -
W^\dagger_{\mu}V_{\nu}W^{\mu\nu}) \\  
& {} + i \kappa_V W^{\dagger}_\mu W_\nu
V^{\mu\nu} + \frac{i\lambda_V}{M^2_W} W^\dagger_{\lambda\mu}W^\mu{}_\nu
V^{\nu\lambda} \\
\end{split}
\end{equation}

\noindent where $V = \gamma\text{ or }Z$, $W^\mu$ is the $W^-$ field,
$W_{\mu\nu} = \partial_\mu W_\nu - \partial_\nu W_\mu$, $V_{\mu\nu} =
\partial_\mu V_\nu - \partial_\nu V_\mu$, and $g_1^\gamma = 1$. The overall couplings are
$g_{WW\gamma} = -e$ and $g_{WWZ} = -e\cot\theta_W$.

The five remaining parameters are
$g^Z_1$, $\kappa_Z$, $\kappa_\gamma$, $\lambda_Z$, and $\lambda_\gamma$.
In the SM, $g^Z_1 = \kappa_Z = \kappa_\gamma = 1$ and $\lambda_Z =
\lambda_\gamma = 0$. The couplings $g^Z_1$, $\kappa_Z$, and $\kappa_\gamma$ are often
written in terms of their deviation from the SM values as
$\Delta g^Z_1 = g^Z_1 - 1$, and similarly for $\Delta \kappa_Z$ and $\Delta \kappa_\gamma$.

One effect of introducing anomalous coupling parameters into the
SM Lagrangian is an increase of the cross section for the
$q\bar{q} \to Z/\gamma \to W^+W^-$ process, particularly as
parton center-of-mass energies rise to infinity.  Thus, constant 
finite values of the anomalous couplings produce unphysically large 
cross sections, violating unitarity. To keep the
cross section from diverging, the anomalous coupling must vanish as $s
\to \infty$. This is done by introducing a dipole form factor for an 
arbitrary coupling $\alpha$ ($g^Z_1$, $\kappa_V$, or $\lambda_V$ from Eq.~\ref{eq-lag}):

\begin{equation}
\alpha(\hat{s}) = \frac{\alpha_0}{(1 + \hat{s}/\Lambda^2)^2} 
\end{equation}

\noindent where the form factor scale $\Lambda$ is set by new physics. For a given
value of $\Lambda$, there is an upper limit on the size of the coupling,
beyond which unitarity is exceeded.

Limits on the $WW\gamma$ and $WWZ$ anomalous
couplings are set using the data, event selection, and background calculations
from the recent $WW$ cross section analysis published by the D\O\ Collaboration
\cite{Abazov:2004kc}. The cross section analysis measures the $p\bar{p} \to WW$ cross
section to be $13.8^{+4.3}_{-3.8}\textrm{(stat)}^{+1.2}_{-0.9}
\textrm{(syst)}\pm0.9\textrm{(lum)} \textrm{ pb}$, compared with a
SM next-to-leading order prediction of $13.0-13.5 \textrm{ pb}$ \cite{Campbell:1999}. 

The leptonic channels $WW \to \ell^+\nu\ell^-\bar{\nu}\ (\ell = e,\ \mu)$ were
used to measure the cross section, with
integrated luminosities of $252\ \mathrm{pb}^{-1}$ for the
$e^+e^-$ channel, $235\ \mathrm{pb}^{-1}$ for the
$e^\pm\mu^\mp$ channel, and $224\ \mathrm{pb}^{-1}$ for the
$\mu^+\mu^-$ channel.  
Table~\ref{tab-cand} summarizes the predicted numbers of signal and background events
and the number of observed candidate events in each channel.
Details of selection cuts and efficiencies can be found in Ref.~\cite{Abazov:2004kc}.

\begin{table}
\begin{tabular}{cr@{$\,\pm \,$}lr@{$\,\pm \,$}lc}
\hline \hline
Channel & \multicolumn{2}{c}{Signal} & \multicolumn{2}{c}{Background} & Candidates \\
\hline 
$e^+e^-$       & 3.26 & 0.05 & 2.30 & 0.21 & 6  \\ 
$e^\pm\mu^\mp$ & 10.8 & 0.1  & 3.81 & 0.17 & 15 \\ 
$\mu^+\mu^-$   & 2.01 & 0.05 & 1.94 & 0.41 & 4  \\ 
\hline \hline
\end{tabular}
\caption{\label{tab-cand}
Predicted numbers of signal and background events, with statistical error, and number of
candidate events for each decay channel.
}
\end{table}

Four anomalous coupling relationships are considered.  In the
first relationship, the $WW\gamma$
parameters are equal to the $WWZ$ parameters: 
$\Delta\kappa_\gamma = \Delta\kappa_Z$ and $\lambda_\gamma = \lambda_Z$.  The
second relationship, the HISZ parameterization \cite{Hagiwara:1993ck}, imposes SU(2)$\times$U(1) 
symmetry upon the coupling parameters.  For the final two relationships,
either the SM $WW\gamma$ or $WWZ$ interaction is fixed, while
the other parameters are allowed to vary. 
In all cases, parameters which are not constrained by the coupling
relationships are set to their SM values.

Anomalous coupling limits must be set for a given coupling relationship
and form factor scale.  Setting limits on a pair of
anomalous couplings simultaneously requires a grid of Monte Carlo (MC)
events, generated specifically for that coupling relationship and form factor scale.
The likelihood of getting the actual measured events
is calculated at each of the grid points and the limits for
the couplings are then extracted from a fit to the likelihood
distribution across the grid.

The leading order MC generator by
Hagiwara, Woodside, and Zeppenfeld (HWZ) \cite{Hagiwara:1989mx} is used
to generate events for a grid in $(\Delta\kappa,\lambda)$ space. 
The central area of each grid has a finer spacing of generated
coupling parameters to ensure that the likelihood surface is well defined 
inside the area where limits are expected to be set.

\begin{figure}
\begin{center}
\includegraphics[width=3.25in]{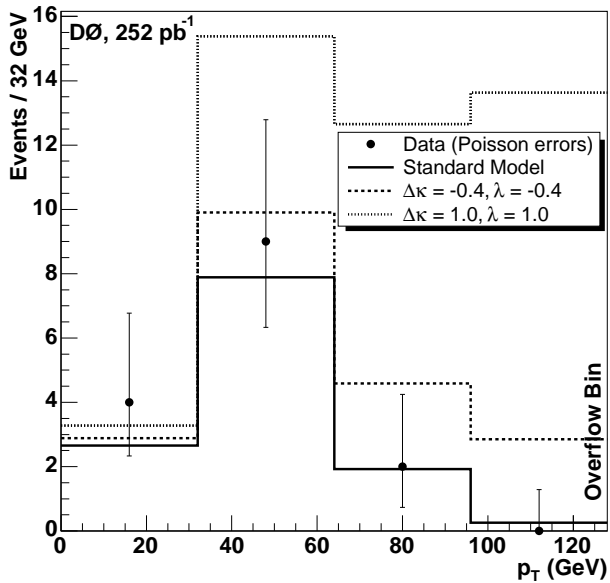} %% old with = 6.5cm
\caption{\label{fig-pt}
Leading lepton $p_T$ distributions for data (points), SM MC (solid line),
and two anomalous coupling MC scenarios (dashed lines), 
from the $WW\to e^\pm \mu^\mp$ channel, binned as used to calculate likelihood.  }
\end{center}
\end{figure}

The generated events for each grid point are passed through 
a parameterized simulation of the D\O\ detector that is
tuned using $Z$ boson events. 
The outputs for each grid point are the simulated $p_T$ spectra for the
two leptons in the event scaled to match the luminosity of the
data. Eight $p_T$ bins are used to calculate the likelihood at each 
grid point: three bins plus
an overflow bin for each of the two leptons.  Figure~\ref{fig-pt} shows
the data for the leading lepton in the $e^\pm \mu^\mp$ channel with MC
estimations for the SM and two sample anomalous coupling grid points.

The simulated signal from the HWZ generator and the background, 
taken from the cross section analysis, are compared to
the $p_T$ distribution of the data by calculating a bin-by-bin
likelihood. Each bin is assumed to have a Poisson distribution with a
mean equal to the sum of the signal and background. The uncertainties on
the signal and background distributions are accounted for by weighting
with Gaussian distributions. Correlations between the signal and
background uncertainties for each channel are small, so they
are handled separately. The uncertainty on the luminosity is 100\%
correlated, and so varies the same way for all channels. The
likelihood, $L$, is calculated as

\begin{equation}
L = \int\mathcal{G}_{f_l} P_{ee}(f_l)P_{e\mu}(f_l)P_{\mu\mu}(f_l)df_l
\end{equation}
\begin{equation}
\begin{split}
P_{\ell\ell}(f_l) = \int\mathcal{G}_{f_n} \int
\mathcal{G}_{f_b}\prod_{i=1}^{N_{\mathrm{bins}}}
\mathcal{P} \left[N_{\ell\ell}^i; \left(f_l f_n n_{\ell\ell}^i \right.\right.\\
\left.\left.+ f_l f_b b_{\ell\ell}^i\right) \right]df_n df_b
\end{split}
\end{equation}

\noindent where $\mathcal{P}(a;\alpha)$ is
the Poisson probability of obtaining $a$ events if the mean expected
number is $\alpha$; $n_{\ell\ell'}^i$ and $b_{\ell\ell'}^i$ are the
simulated numbers of signal and background events for the $\ell\ell'$
channel in bin $i$; $N_{\ell\ell'}^i$ is the measured number of events
for this channel in this bin; and $f_l$, $f_n$, and $f_b$ are the
luminosity, signal, and background weights drawn from the Gaussian
distributions $\mathcal{G}_{f_l}$, $\mathcal{G}_{f_n}$, and
$\mathcal{G}_{f_b}$ respectively.

To extract the limits, a 6th order polynomial is fitted to the grid of
negative log likelihood values.  
The one- and two-dimensional 95\% C.L. limits are
determined by integrating the likelihood curve or surface, respectively.
In the one-dimensional case, the 95\% C.L. limits represent the
pair of points of equal likelihood that bound 95\% of the total
integrated area between the ends of the MC grid.  The two-dimensional
95\% C.L. countour line is the set of points of equal likelihood
that bound a region containing 95\% of the total integrated volume
between the MC grid boundaries.

\begin{table}[!tb]
\begin{tabular}{l|c|c|c}
\hline \hline
\multicolumn{2}{c|}{Coupling} &
\multicolumn{1}{c|}{95\% C.L. Limits} &
\multicolumn{1}{c}{$\Lambda$ (TeV)} 
\\ \hline 
& $\lambda $ & $-0.31, 0.33$ & \\
\raisebox{1.5ex}[0pt]{$WW\gamma = WWZ$} & $\Delta\kappa $ & $-0.36, 0.47$ & \raisebox{1.5ex}[0pt]{1.5} \\ \hline
& $\lambda$ & $-0.29, 0.30$ &  \\
\raisebox{1.5ex}[0pt]{$WW\gamma = WWZ$} & $\Delta\kappa $ & $-0.32, 0.45$ & \raisebox{1.5ex}[0pt]{2.0} \\ \hline
& $\lambda $ & $-0.34, 0.35$ &  \\ 
\raisebox{1.5ex}[0pt]{HISZ} & $\Delta\kappa_\gamma$ & $-0.57, 0.75$ & \raisebox{1.5ex}[0pt]{1.5} \\ \hline
& $\lambda_Z$ & $-0.39, 0.39$ &  \\ 
\raisebox{1.5ex}[0pt]{SM $WW\gamma $} & $\Delta\kappa_Z$ & $-0.45, 0.55$ & \raisebox{1.5ex}[0pt]{2.0} \\ \hline
& $\lambda_\gamma$ & $-0.97, 1.04$ &  \\ 
\raisebox{1.5ex}[0pt]{SM $WWZ$} & $\Delta\kappa_\gamma$ & $-1.05, 1.29$ & \raisebox{1.5ex}[0pt]{1.0} \\ 
\hline \hline
\end{tabular}
\caption{\label{tab-limits}
One-dimensional limits
at the 95\% C.L. with various assumptions relating the 
$WW\gamma$ and $WWZ$ couplings at various values of
$\Lambda$. Parameters which are not constrained by the coupling
relationships are set to their SM values.
}
\end{table}

\begin{figure}
\begin{center}
\includegraphics[width=3.25in]{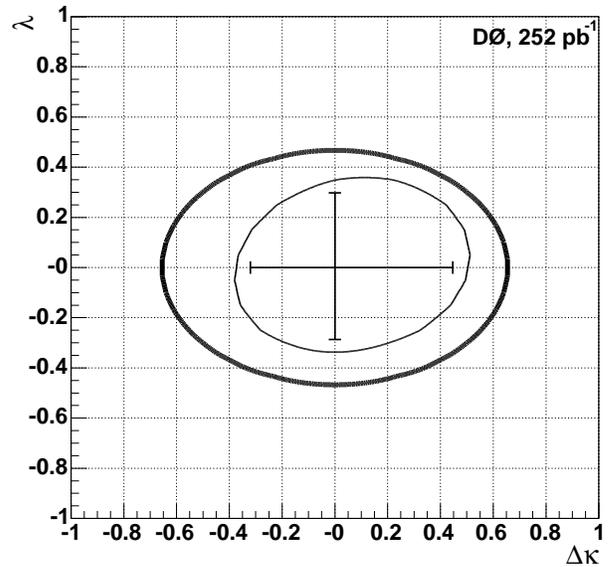} %% old with = 6.5cm
\caption{\label{fig-Zeqg2.0}
One- and two-dimensional 95\% C.L. limits when $WWZ$ couplings are
equal to $WW\gamma$ couplings, at $\Lambda=2.0$ TeV.  The bold curve
is the unitarity limit, the inner curve is the two-dimensional 95\% C.L. contour,
and the ticks along the axes are the one-dimensional 95\% C.L. limits.
}
\end{center}
\end{figure}

\begin{figure}
\begin{center}
\mbox{
\includegraphics[width=1.66in]{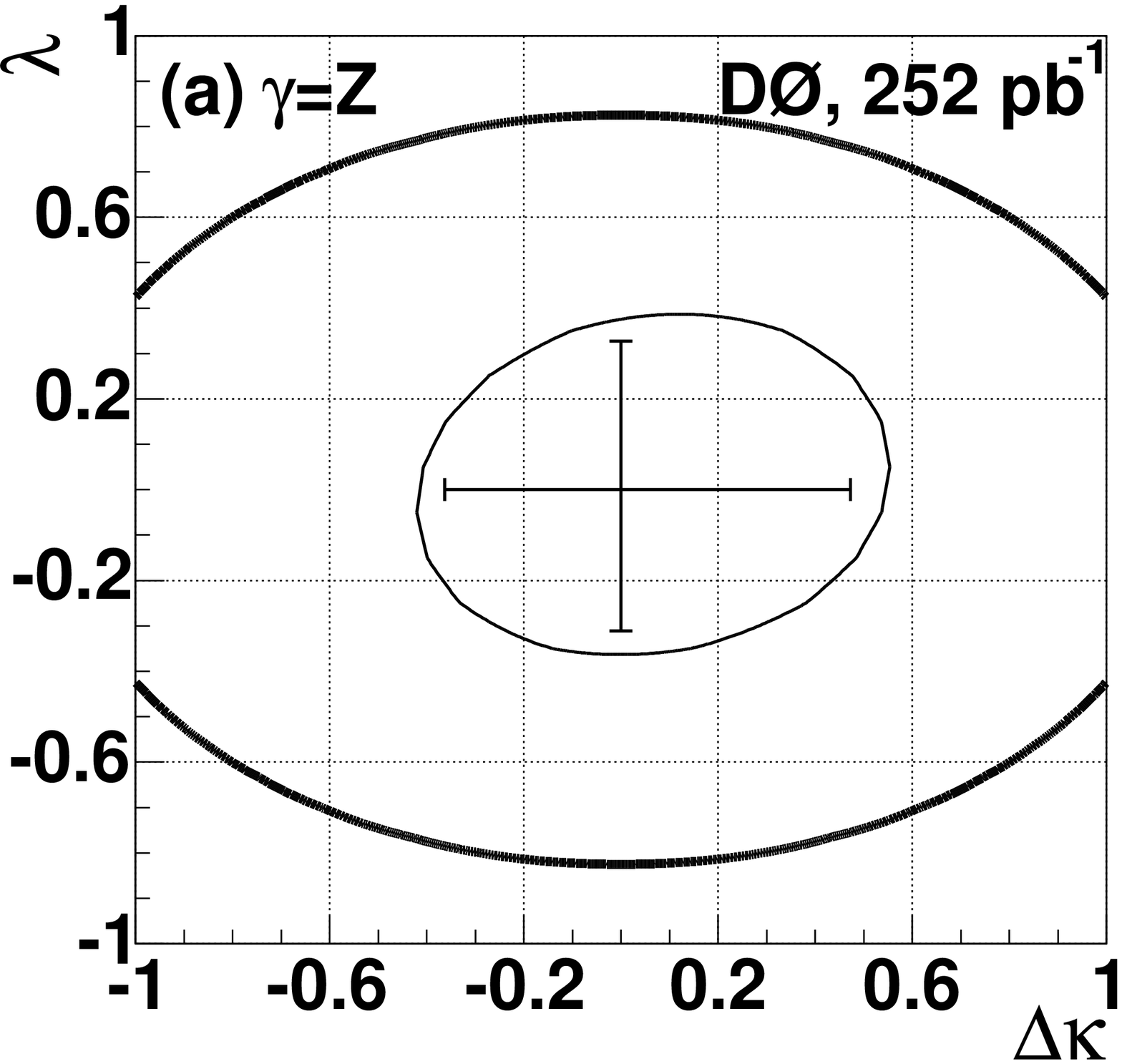}\hfill % Old value: 4.2cm
\includegraphics[width=1.66in]{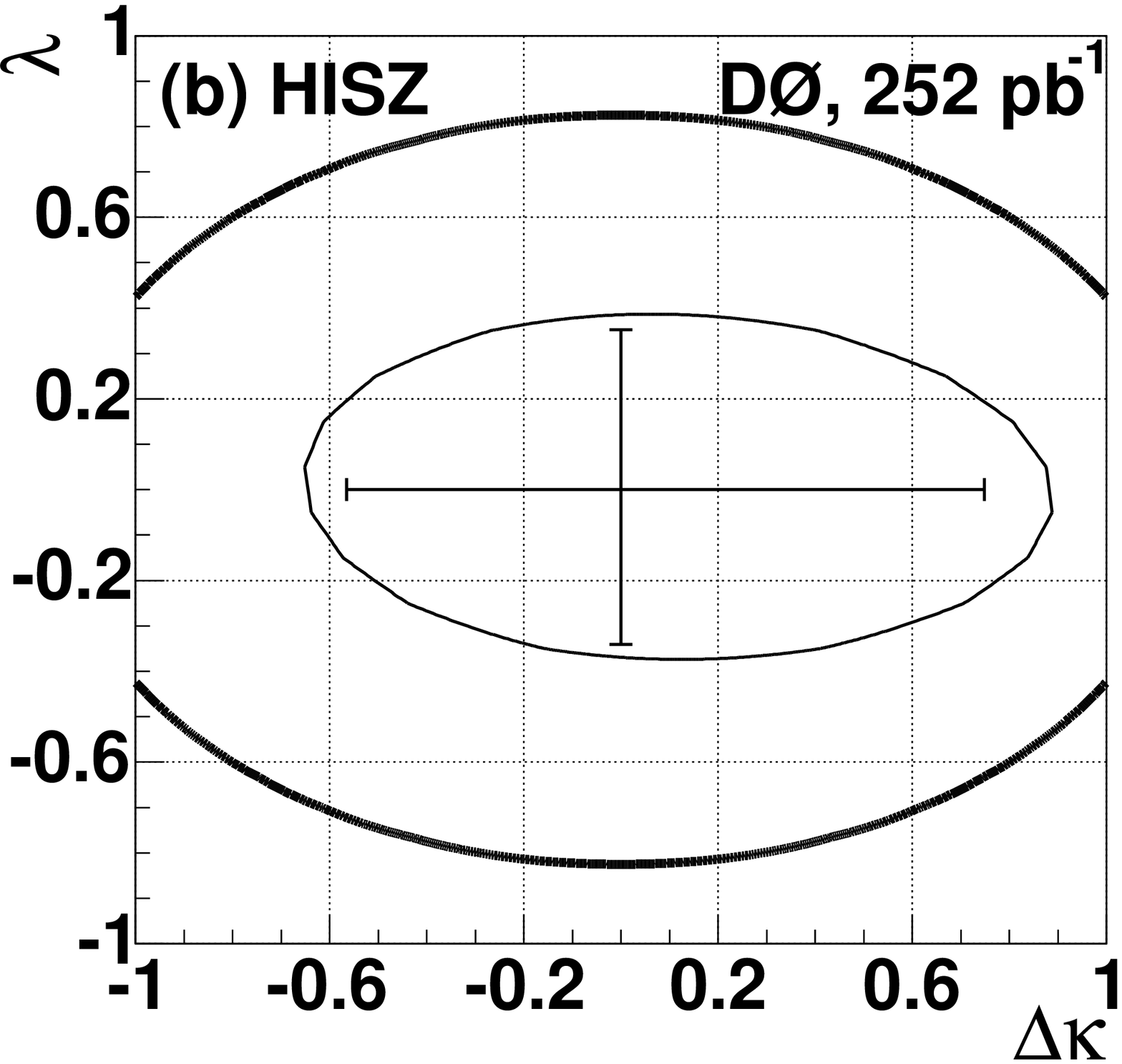} 
}
\mbox{
\includegraphics[width=1.66in]{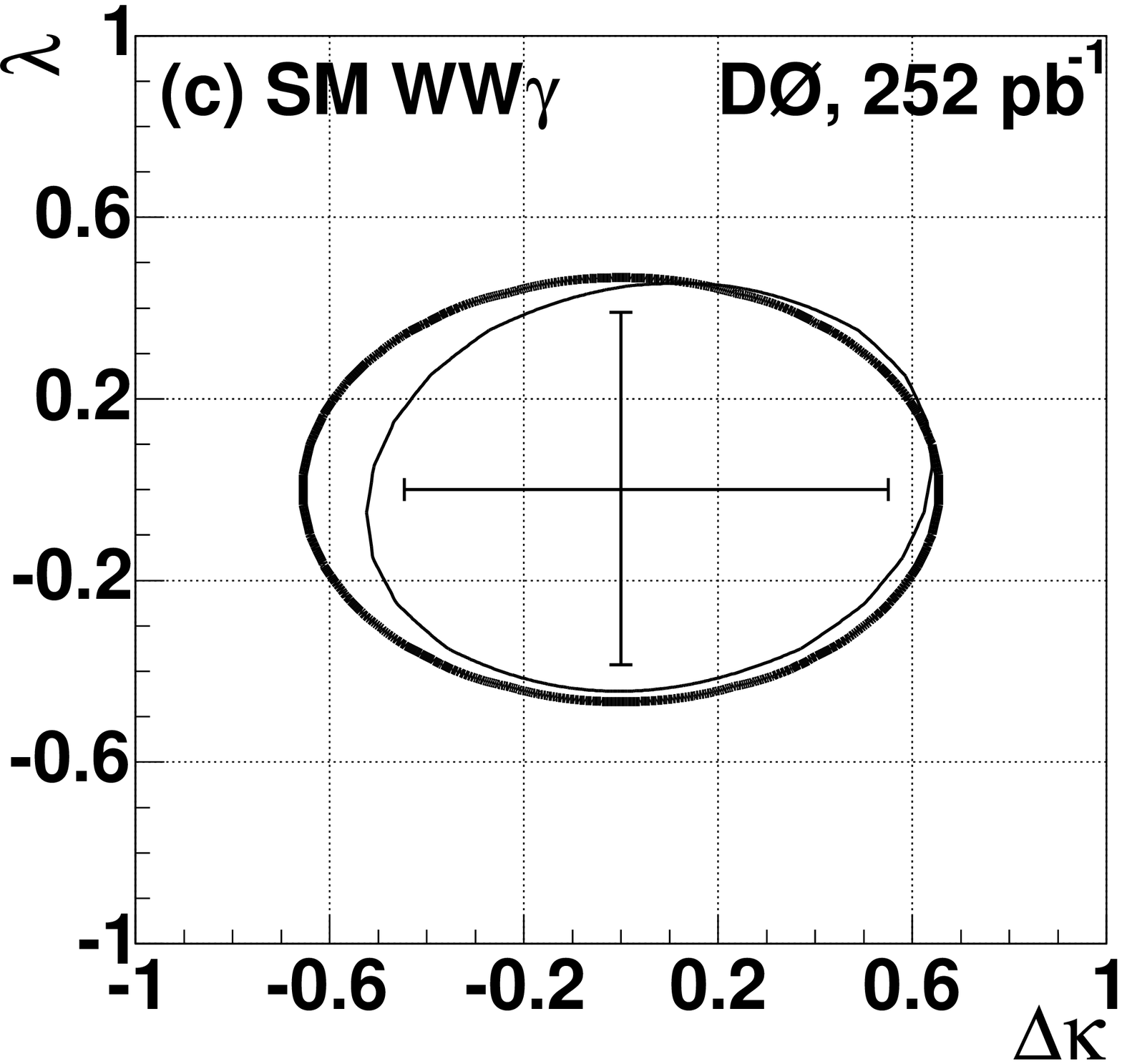}\hfill
\includegraphics[width=1.66in]{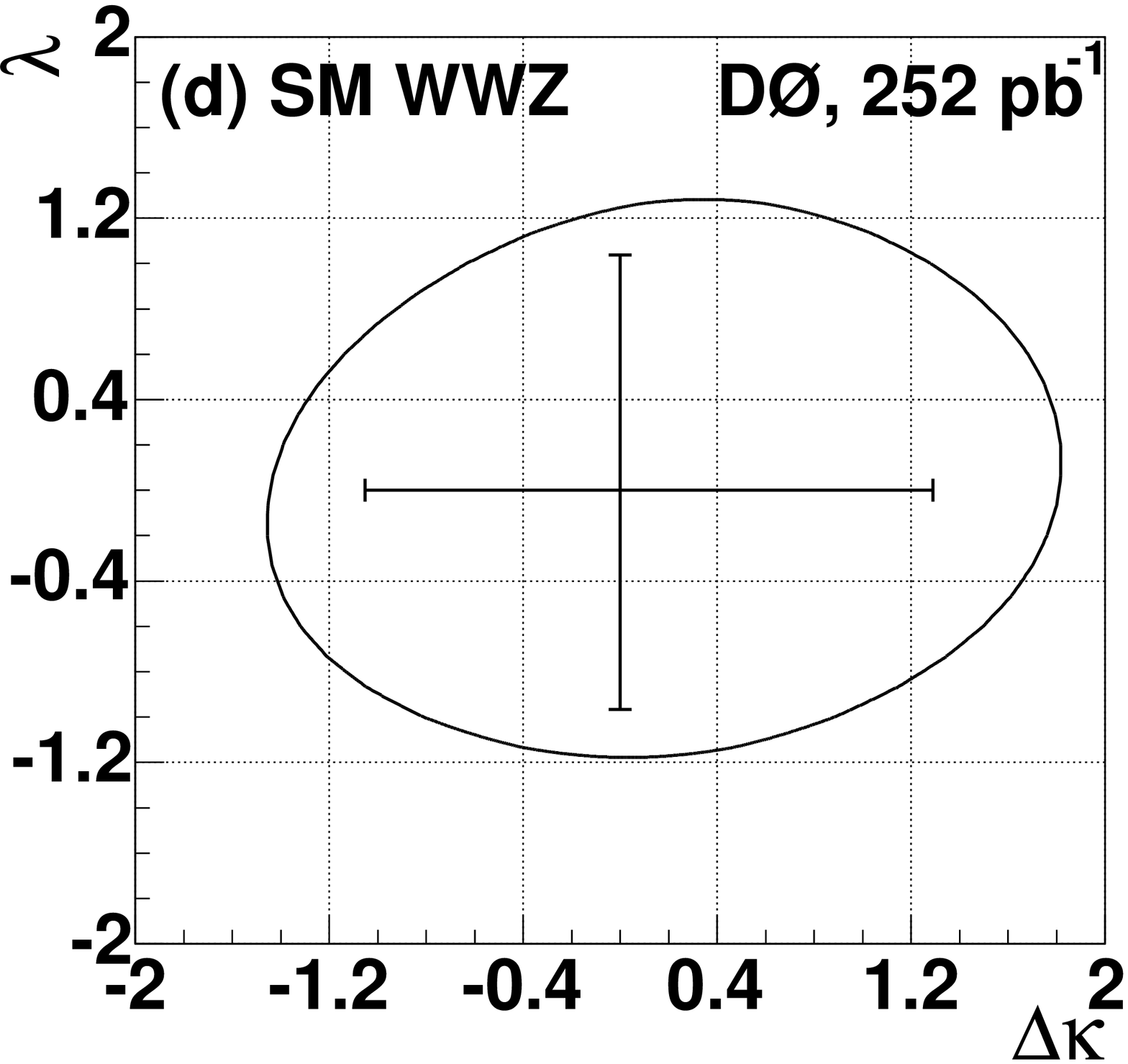}
}
\caption{\label{fig-other}
One- and two-dimensional 95\% C.L. limits for (a) $WW\gamma=WWZ$ at $\Lambda$=1.5 TeV,
(b) HISZ at $\Lambda$=1.5 TeV, (c) SM $WW\gamma$ at $\Lambda$=2.0 TeV, 
and (d) SM $WWZ$ at $\Lambda$=1.0 TeV. 
The bold curve is the unitarity limit (where it fits within the plot
boundaries), the inner curve is the two-dimensional 95\% C.L. contour,
and the ticks along the axes are the one-dimensional 95\% C.L. limits.
}
\end{center}
\end{figure}

One-dimensional 95\% C.L. limits are summarized in 
Table~\ref{tab-limits}, and two-dimensional 95\% C.L. contours are shown in 
Figs.~\ref{fig-Zeqg2.0} and \ref{fig-other}.
Under the assumption that the $WW\gamma$ and $WWZ$ couplings are equal
and using a form factor scale of $\Lambda = 2.0\text{ TeV}$,
the 95\% C.L. limits obtained are $-0.32 < \Delta\kappa < 0.45$ and $-0.29 < \lambda < 0.30$.
This significantly improves upon the previous limits from the D\O\ Collaboration,
$-0.62 < \Delta\kappa < 0.77$ and $-0.53 < \lambda < 0.56$, set
in Run I at the Fermilab Tevatron Collider for the same channels under the same assumption using 
an integrated luminosity of $100 \text{ pb}^{-1}$ \cite{Abbott:1998gp}.
Although the combined anomalous coupling limits from the 
CERN $e^+e^-$ Collider (LEP) collaborations are tighter \cite{LEP:2005}, 
the hadronic collisions at the Fermilab Tevatron Collider 
explore a range of parton center-of-mass energies not explored at LEP.

\begin{acknowledgments}
% acknowledgement_paragraph_r2.tex                7/19/06
%
We thank the staffs at Fermilab and collaborating institutions, 
and acknowledge support from the 
DOE and NSF (USA);
CEA and CNRS/IN2P3 (France);
FASI, Rosatom and RFBR (Russia);
CAPES, CNPq, FAPERJ, FAPESP and FUNDUNESP (Brazil);
DAE and DST (India);
Colciencias (Colombia);
CONACyT (Mexico);
KRF and KOSEF (Korea);
CONICET and UBACyT (Argentina);
FOM (The Netherlands);
PPARC (United Kingdom);
MSMT (Czech Republic);
CRC Program, CFI, NSERC and WestGrid Project (Canada);
BMBF and DFG (Germany);
SFI (Ireland);
The Swedish Research Council (Sweden);
Research Corporation;
Alexander von Humboldt Foundation;
and the Marie Curie Program.

\end{acknowledgments}

%\bibliographystyle{h-physrev}
\bibliography{wwacpaper4.bib}

\begin{thebibliography}{6}
\expandafter\ifx\csname natexlab\endcsname\relax\def\natexlab#1{#1}\fi
\expandafter\ifx\csname bibnamefont\endcsname\relax
  \def\bibnamefont#1{#1}\fi
\expandafter\ifx\csname bibfnamefont\endcsname\relax
  \def\bibfnamefont#1{#1}\fi
\expandafter\ifx\csname citenamefont\endcsname\relax
  \def\citenamefont#1{#1}\fi
\expandafter\ifx\csname url\endcsname\relax
  \def\url#1{\texttt{#1}}\fi
\expandafter\ifx\csname urlprefix\endcsname\relax\def\urlprefix{URL }\fi
\providecommand{\bibinfo}[2]{#2}
\providecommand{\eprint}[2][]{\url{#2}}

\bibitem[{\citenamefont{Hagiwara et~al.}(1990)\citenamefont{Hagiwara, Woodside,
  and Zeppenfeld}}]{Hagiwara:1989mx}
\bibinfo{author}{\bibfnamefont{K.}~\bibnamefont{Hagiwara}},
  \bibinfo{author}{\bibfnamefont{J.}~\bibnamefont{Woodside}}, \bibnamefont{and}
  \bibinfo{author}{\bibfnamefont{D.}~\bibnamefont{Zeppenfeld}},
  \bibinfo{journal}{Phys. Rev. D} \textbf{\bibinfo{volume}{41}},
  \bibinfo{pages}{2113} (\bibinfo{year}{1990}).

\bibitem[{\citenamefont{Abazov et~al.}(2005)}]{Abazov:2004kc}
\bibinfo{author}{\bibfnamefont{V.~M.} \bibnamefont{Abazov}}
  \emph{et~al.} (\bibinfo{collaboration}{D\O\ Collaboration}),
  \bibinfo{journal}{Phys. Rev. Lett.} \textbf{\bibinfo{volume}{94}},
  \bibinfo{pages}{151801}, \eprint{hep-ex/0410066} (\bibinfo{year}{2005}).

\bibitem[{\citenamefont{Campbell and Ellis}(1999)}]{Campbell:1999}
\bibinfo{author}{\bibfnamefont{J.~M.} \bibnamefont{Campbell}} \bibnamefont{and}
  \bibinfo{author}{\bibfnamefont{R.~K.} \bibnamefont{Ellis}},
  \bibinfo{journal}{Phys. Rev. D} \textbf{\bibinfo{volume}{60}},
  \bibinfo{pages}{113006}, \eprint{hep-ph/9905386} (\bibinfo{year}{1999}).

\bibitem[{\citenamefont{Hagiwara et~al.}(1993)\citenamefont{Hagiwara, Ishihara,
  Szalapski, and Zeppenfeld}}]{Hagiwara:1993ck}
\bibinfo{author}{\bibfnamefont{K.}~\bibnamefont{Hagiwara}},
  \bibinfo{author}{\bibfnamefont{S.}~\bibnamefont{Ishihara}},
  \bibinfo{author}{\bibfnamefont{R.}~\bibnamefont{Szalapski}},
  \bibnamefont{and}
  \bibinfo{author}{\bibfnamefont{D.}~\bibnamefont{Zeppenfeld}},
  \bibinfo{journal}{Phys. Rev. D} \textbf{\bibinfo{volume}{48}},
  \bibinfo{pages}{2182} (\bibinfo{year}{1993}), \bibinfo{note}{the coupling
  relationships used are $\Delta\kappa_Z = \Delta\kappa_\gamma (1 - \tan^2
  \theta_W)$, $\Delta g_1^Z = \Delta\kappa_\gamma / (2 \cos^2 \theta_W)$, and
  $\lambda_Z = \lambda_\gamma$}.

\bibitem[{\citenamefont{Abbott et~al.}(1998)}]{Abbott:1998gp}
\bibinfo{author}{\bibfnamefont{B.}~\bibnamefont{Abbott}} \emph{et~al.}
  (\bibinfo{collaboration}{D\O\ Collaboration}), \bibinfo{journal}{Phys. Rev.
  D} \textbf{\bibinfo{volume}{58}}, \bibinfo{pages}{031102}, \eprint{hep-ex/9803017}
  (\bibinfo{year}{1998}).

\bibitem[{\citenamefont{{\relax The LEP Collaborations ALEPH}
  et~al.}(2005)\citenamefont{{\relax The LEP Collaborations ALEPH}, DELPHI, L3,
  OPAL, and the LEP TGC Working~Group}}]{LEP:2005}
\bibinfo{author}{\bibnamefont{{\relax The LEP Collaborations ALEPH}}},
  \bibinfo{author}{\bibnamefont{DELPHI}}, \bibinfo{author}{\bibnamefont{L3}},
  \bibinfo{author}{\bibnamefont{OPAL}}, \bibnamefont{and}
  \bibinfo{author}{\bibnamefont{the LEP TGC Working~Group}},
  \bibinfo{journal}{LEPEWWG/TGC/2005-01}  (\bibinfo{year}{2005}).

\end{thebibliography}

\end{document}